\documentclass[preprint,showpacs,preprintnumbers,amsmath,amssymb]{revtex4}

\usepackage{graphicx}
\usepackage{dcolumn}
\usepackage{bm}

\begin{document}
\title{Unexpected drop of dynamical heterogeneities in
colloidal suspensions approaching the jamming transition}
\author{Pierre Ballesta$^{1,2}$, Agn\`{e}s Duri$^{1,3}$, and Luca Cipelletti$^{1*}$}
\affiliation{$^{1}$LCVN UMR 5587, Universit\'{e} Montpellier 2 and
CNRS, 34095 Montpellier Cedex 5, France
\\ $^2$ Present Address:
HASYLAB at DESY, D-22603 Hambourg, Germany
\\ $^3$ Present address: School of Physics,
The University of Edinburgh, Edinburgh EH9 3JZ, UK}

\email{lucacip@lcvn.univ-montp2.fr}
\date{\today}

\begin{abstract}

\textbf{As the glass (in molecular fluids\cite{Donth}) or the
jamming (in colloids and grains\cite{LiuNature1998}) transitions
are approached, the dynamics slow down dramatically with no marked
structural changes. Dynamical heterogeneity (DH) plays a crucial
role: structural relaxation occurs through correlated
rearrangements of particle ``blobs'' of size
$\xi$\cite{WeeksScience2000,DauchotPRL2005,Glotzer,Ediger}. On
approaching these transitions, $\xi$ grows in
glass-formers\cite{Glotzer,Ediger},
colloids\cite{WeeksScience2000,BerthierScience2005}, and driven
granular materials\cite{KeysNaturePhys2007} alike, strengthening
the analogies between the glass and the jamming transitions.
However, little is known yet on the behavior of DH very close to
dynamical arrest. Here, we measure in colloids the maximum of a
``dynamical susceptibility'', $\chi^*$, whose growth is usually
associated to that of $\xi$\cite{LacevicPRE}. $\chi^*$ initially
increases with volume fraction $\varphi$, as
in\cite{KeysNaturePhys2007}, but strikingly drops dramatically
very close to jamming. We show that this unexpected behavior
results from the competition between the growth of $\xi$ and the
reduced particle displacements associated with rearrangements in
very dense suspensions, unveiling a richer-than-expected
scenario.}
\end{abstract}


\maketitle

The recent observation of a critical-like growth of temporal and
spatial dynamical fluctuations in a 2D athermal system approaching
jamming\cite{KeysNaturePhys2007}, similar to that hypothesized for
glass formers\cite{Biroli}, has raised hope that the glass and the
jamming transition may be unified, calling at the same time for
further, tighter experimental verifications. Here, we investigate
temporal DH in a 3D thermal system, concentrated colloidal
suspensions close to the maximum packing fraction. Temporal and
spatial DH are usually closely related: the former can be
quantified by a ``four-point dynamical susceptibility'' $\chi_4$
(the variance of a time-resolved correlation function describing
the system relaxation), whose amplitude is proportional to
$\xi^3$\cite{somerefssimul,LacevicPRE,JP}. Surprisingly, we find
that very close to jamming temporal and spatial DH decouple: while
$\xi$ continuously grows with volume fraction, the amplitude of
temporal fluctuations drops sharply close to the maximum packing
fraction. These findings challenge current scenarios where the
slowing down of the dynamics on approaching jamming is accompanied
by enhanced dynamical fluctuations.

We study concentrated suspensions of polyvinilchloride (PVC)
xenospheres\cite{someREfXenospheres} suspended in dioctilphtalate
(DOP). The particles are highly polydisperse, with typical
diameter $\approx 10$ $\mu$m; they behave as slightly deformable
hard spheres. The samples are loaded in cells of thickness $L=2$
mm, vigorously stirred and outgassed to remove air bubbles. The
dynamics are probed by dynamic light scattering in the highly
multiple scattering limit (Diffusing Wave Spectroscopy,
DWS\cite{DWSGeneral}), adopting the transmission geometry, with
$L/\ell^* \approx 10$, where $\ell^*$ is the photon transport mean
path. 
This technique allows the dynamics to be probed on length scales
as small as a few nm\cite{DWSGeneral}, which match well the
restrained motion in tightly packed suspensions. A
charge-coupled-device (CCD) detector is used to record the speckle
pattern scattered by the sample. The evolution of the speckle
images is quantified by the two-time degree of
correlation\cite{DuriPRE2005} $c_I(q,t,\tau) = \left<
I_p(t)I_p(t+\tau) \right >_p / \left( \left <I_p(t) \right
>_p \left <I_p(t + \tau ) \right >_p \right) - 1$, where $I_p(t)$
is the scattered intensity at pixel $p$ and time $t$ and $\left <
\cdot \cdot \cdot \right>_p$ is an average over the CCD pixels.

In order to follow the evolution of the dynamics, we calculate
$g_2(t,\tau)-1$, the two-time intensity correlation function
obtained by averaging $c_I(t,\tau)$ over a few CCD frames. Due to
the limited acquisition rate of the CCD, the initial decay of
$g_2(t,\tau)-1$ is not captured; at longer delays, a plateau
followed by a final relaxation is observed (see Fig. 2a for an
example of a time-averaged $g_2$), indicative of very slow
rearrangements. Figure 1a shows a typical example of the time
dependence of $\tau_0$, the characteristic time of the final
relaxation obtained by fitting $g_2-1$ to a stretched exponential
$a(t) \exp\left\{-\left[\tau/\tau_0(t)\right]^{\beta(t)}\right\}$.
Initially, $\tau_0$ grows nearly linearly with $t$, as observed in
many glassy systems. However, for $t > 28000$ sec a stationary
regime is observed, where $\tau_0$ exhibits surprisingly large
fluctuations but no overall increasing trend. A similar behavior
is observed for all volume fractions; all data presented in the
following refer to the stationary regime\cite{noteduration}. We
first investigate the $\varphi$-dependence of the average
dynamics, as quantified by $\overline{\tau_0}$ and
$\overline{\beta}$, where $\overline{\cdot\cdot\cdot}$ denotes a
time average. As shown in Fig. 1b, $\overline{\tau_0}$
continuously increases with $\varphi$. Data taken for freshly
prepared samples (solid circles) can be fitted by a critical law,
$\overline{\tau_0} \sim 1/| \varphi/\varphi_{\rm max} - 1 |^{x}$,
with $\varphi_{\rm max} = 0.752$, consistent with expectations for
highly polydisperse samples\cite{packing}, and $x = 1.01 \pm
0.04$, similarly to Ref.\cite{KeysNaturePhys2007}. We also study
samples that have been aged for several days and whose dynamics is
re-initialized by vigorously stirring and outgassing them
(semi-open circles). Their effective volume fraction is higher
than the nominal one, due to the slight swelling of PVC particles
suspended in DOP for very long times\cite{someREfXenospheres},
allowing to achieve an even tighter packing. To compare the
dynamics of both fresh and aged samples, we assign an effective
volume fraction, $\varphi_{\rm eff}$, to the latter so that their
average relaxation time falls on the critical-like curve
determined for the fresh samples. Figure 1c shows
$\overline{\beta}(\varphi)$: at the lowest volume fraction, the
shape of $g_2$ is slightly stretched ($\overline{\beta} = 0.86 <
1$), similarly to what observed for correlation functions in many
glassy systems\cite{Donth}. Surprisingly, as $\varphi$ increases
$\overline{\beta}$ grows above one, finally saturating around
$1.3$. A similar ``compressed'' exponential relaxation has been
observed in single \cite{LucaPRL,LaponiteXPCS,ReviewLeheney} and
multiple\cite{Munch,Leheny} scattering experiments on systems
close to jamming, usually associated with ultra-slow ballistic
motion.

We quantify the temporal fluctuations of the dynamics by
calculating $\chi(\tau,\varphi)$, the relative variance of $c_I$,
defined by
\begin{equation}
\chi(\tau) \equiv \chi(\tau,\varphi) =
\overline{\left(c_I(t,\tau)- \overline{c_I(t,\tau)}\right ) ^2 } /
\overline{a}^{\,2} \, ,
   \label{Eq:chi}
\end{equation}
where the normalization is introduced to account for the
$\varphi$-dependence of the amplitude of the final relaxation of
$g_2-1$; data are furthermore corrected for experimental
noise\cite{DuriPRE2005}. The variance introduced above corresponds
to the dynamical susceptibility $\chi_4$ much studied in
simulations of glass formers
\cite{notenorm,LacevicPRE,somerefssimul}. Figure 2a shows both the
average correlation function $\overline{g_2}-1$ (open circles) and
$\chi$ (solid circles) for $\varphi = 0.738$. The dynamical
susceptibility exhibits a marked peak around $\tau_0$, a direct
manifestation of DH also found in many other glassy
systems\cite{LacevicPRE,somerefssimul,MayerPRL2004,DauchotPRL2005,JP,KeysNaturePhys2007}.

Figure 2b shows the height of the peak of the dynamical
susceptibility, $\chi^*$, as a function of $\varphi$. At the
lowest volume fractions, $\chi^*$ increases with $\varphi$; the
data can be fitted by a critical law $\chi^* \sim 1/|
\varphi/\varphi_{\rm max} - 1 |^{y}$ with $y=1.5 \pm 0.2$ (line in
Fig. 2b), close to $y=1.70$ recently reported for driven
grains\cite{KeysNaturePhys2007}. This growing trend is also
analogous to that observed in simulations of glass
formers\cite{Glotzer} and
colloids\cite{WeeksScience2000,BerthierScience2005} (albeit at
lower $\varphi$) and has been interpreted as due to a growing
dynamical length scale on approaching dynamical arrest. At higher
volume fractions, however, an opposite trend is observed: the
amplitude of dynamical fluctuations dramatically decreases close
to $\varphi_{\rm max}$. This striking behavior represents our
central result, which challenges current views of DH close to
dynamical arrest. The unexpected drop of dynamical fluctuations
very close to jamming is confirmed by the non-monotonic behavior
of the width of the temporal distributions of $\tau_0$ and
$\beta$, shown by the vertical bars in Fig. 1b-c. The dispersion
of both parameters initially increases with $\varphi$, but is
eventually reduced close to $\varphi_{\rm max}$, further
demonstrating reduced DH.

We propose that the non-monotonic behavior of $\chi^*$ results
from a competition between the growth of $\xi$ on approaching
$\varphi_{\rm max}$ and the reduced particle displacement
associated with rearrangement events close to jamming, due to
tighter
packing\cite{Weekscage,DauchotCage,PRLRecentGranular,Wyart}.
Indeed, as $\xi$ increases, fewer statistically independent
dynamical regions are contained in the sample, leading to enhanced
fluctuations\cite{MayerPRL2004}. Conversely, as particle
displacement decreases $\chi^*$ is reduced, since more events are
required to significantly decorrelate the scattered light and
fluctuations on a time scale $\sim \overline{\tau_0}$ tend to be
averaged out. These competing mechanisms should be quite general
and should be observable in a variety of systems, provided that DH
are probed close enough to dynamical arrest.

We have incorporated these ideas in a simple model for DWS for a
dynamically heterogeneous process, significantly extending
previous work on the intermittent dynamics of
foams\cite{DurianScience1992} and gels\cite{DuriEPL2006}. The
dynamics is assumed to be due to discrete rearrangement events of
volume $\xi^3$ that occur randomly in space and time; however, in
contrast to Ref.\cite{DurianScience1992} we assume that several
events will be in general necessary to fully decorrelate the phase
of scattered photons, since in concentrated suspensions the
particle displacement associated with one single event may be much
smaller than the wavelength of the light. The two-time field
correlation function for a photon crossing the cell along a path
of length $s$ may then be written as
\begin{equation}
g_1^{(s)}(t,\tau) = \exp[-n_s(t,\tau)^p\sigma^2_{\phi}] \,.
   \label{Eq:g1s}
\end{equation}
Here, $n_s(t,\tau)$ is the number of events along the path between
time $t$ and $t+\tau$ and $\sigma^2_{\phi}$ is the variance of the
change of phase of a photon due to one single event, related to
the particle mean squared displacement associated with such event,
$\sigma^2$, by $\sigma^2_{\phi} \approx  20 \sigma^2/\mu{\mathrm
m}^2$\cite{notemsd}. For a totally uncorrelated change of photon
phase due to distinct events one has $p=1$, while in the opposite
limit of a perfectly correlated change of phase
$p=2$\cite{DWSGeneral}.

We implement our model in Monte Carlo simulations where photon
paths are random walks on a square lattice with lattice parameter
$\ell^*$. The lattice sites are affected by random rearrangement
events of volume $\xi^3$, occurring at a constant rate per unit
volume. The simulated degree of correlation is calculated from
$c_{I,\mathrm{sim}}(t,\tau) = \left[ N_s^{-1}\sum
g_1^{(s)}(t,\tau) \right]^2$, where the sum is over $N_s = 200000$
photon paths and $g_1^{(s)}$ is calculated according to
(\ref{Eq:g1s}). The two-time intensity correlation function,
$g_{2,\mathrm{sim}}(t,\tau)$, and its fluctuations,
$\chi_{\mathrm{sim}}$, are then calculated from
$c_{I,\mathrm{sim}}$ as for the experiments. We vary the control
parameters in the simulation, $\xi^3$, $p$, and
$\sigma^2_{\phi}$\cite{noterate}, to reproduce the experimental
$\varphi$-dependence of $\overline{\beta}$ and $\chi^*$. Since the
particle displacement resulting from one rearrangement ---and thus
$\sigma^2_{\phi}$--- is expected to decrease as $\varphi$ grows,
due to tighter particle
packing\cite{Weekscage,DauchotCage,PRLRecentGranular,Wyart}, we
choose $1/\sigma^2_{\phi}$ as the control parameter against which
simulation results are presented, corresponding to increasing
volume fractions in Fig.~1b-c). Figure 3a shows $\overline{\beta}$
$vs$ $1/\sigma^2_{\phi}$. The data are obtained using the values
of $\xi^3$ shown in Fig. 3c; however, we find that
$\overline{\beta}$ depends only very weakly on $\xi^3$. For large
particle displacements (small $1/\sigma^2_{\phi}$),
$\overline{\beta} \lesssim 1$ in fair agreement with the
experimental value at the lowest $\varphi$. As particle
displacements become increasingly restrained, $\overline{\beta}$
grows and saturates at $\overline{\beta} \approx 1.3$, close to
the experimental values at the highest $\varphi$. The saturation
value depends on the choice of $p$\cite{notesaturation}: here,
$p=1.65$, showing that the change of phase of a photon due to
distinct rearrangements is partially correlated. It is unlikely
that such a correlation exists for events occurring in
non-overlapping regions; by contrast, successive events in the
same region will lead to partially correlated changes of phase
when the direction of displacement persists during several events.
Thus, $p=1.65$ indicates intermittent supradiffusive motion, a
behavior close to the ballistic motion reported for many jammed
systems\cite{LucaPRL,LaponiteXPCS,ReviewLeheney,Munch,Leheny}.

Figure 3b shows the $1/\sigma^2_{\phi}$ dependence of $\chi^*$.
The simulations reproduce well both the non-monotonic trend and
the absolute values of the experimental $\chi^*$. They reproduce
also the non-monotonic $\varphi$-dependence of the dispersion of
$\beta$ (bars in Fig. 3a), once again matching closely the
experimental data. In spite of the drop of dynamical fluctuations
close to $\varphi_{\rm max}$, $\xi^3$ grows steadily with
$\varphi$ (Fig. 3c), until rearrangement events span the whole
sample, corresponding to $\xi \approx 2000$ particle diameters.
System-spanning ``earthquakes'' have been reported in simulations
of both thermal\cite{Kob2000} and athermal\cite{Wyart} systems
close to dynamical arrest, but have never been observed
experimentally.

Our results show that the behavior of dynamical heterogeneity very
close to the jamming transition is much richer and complex than
expected. On the one hand, the growth of dynamically correlated
regions is limited only by the system size, implying that
confinement effects, usually observed on the scale of tens of
particles at most\cite{ReviewConfinement}, should become relevant
macroscopically. On the other hand, temporal fluctuations of the
dynamics are suppressed, challenging current views of DH close to
jamming and calling for new theories.

\vspace{1 cm}

\textit{Acknowledgments} We thank L. Berthier for illuminating
discussions and M. Clo\^{i}tre for providing us with the samples.
This work was partially supported by the European MCRTN ``Arrested
matter'' (MRTN-CT-2003-504712), the NoE ``SoftComp''
(NMP3-CT-2004-502235), and ACI JC2076 and CNES grants. L.C. is a
junior member of the Institut Universitaire de France, whose
support is gratefully acknowledged.

\vspace{1 cm}

 \textit{Competing financial interests} The authors
declare no competing financial interests.

\newpage

\begin{figure}
\includegraphics[scale=0.8]{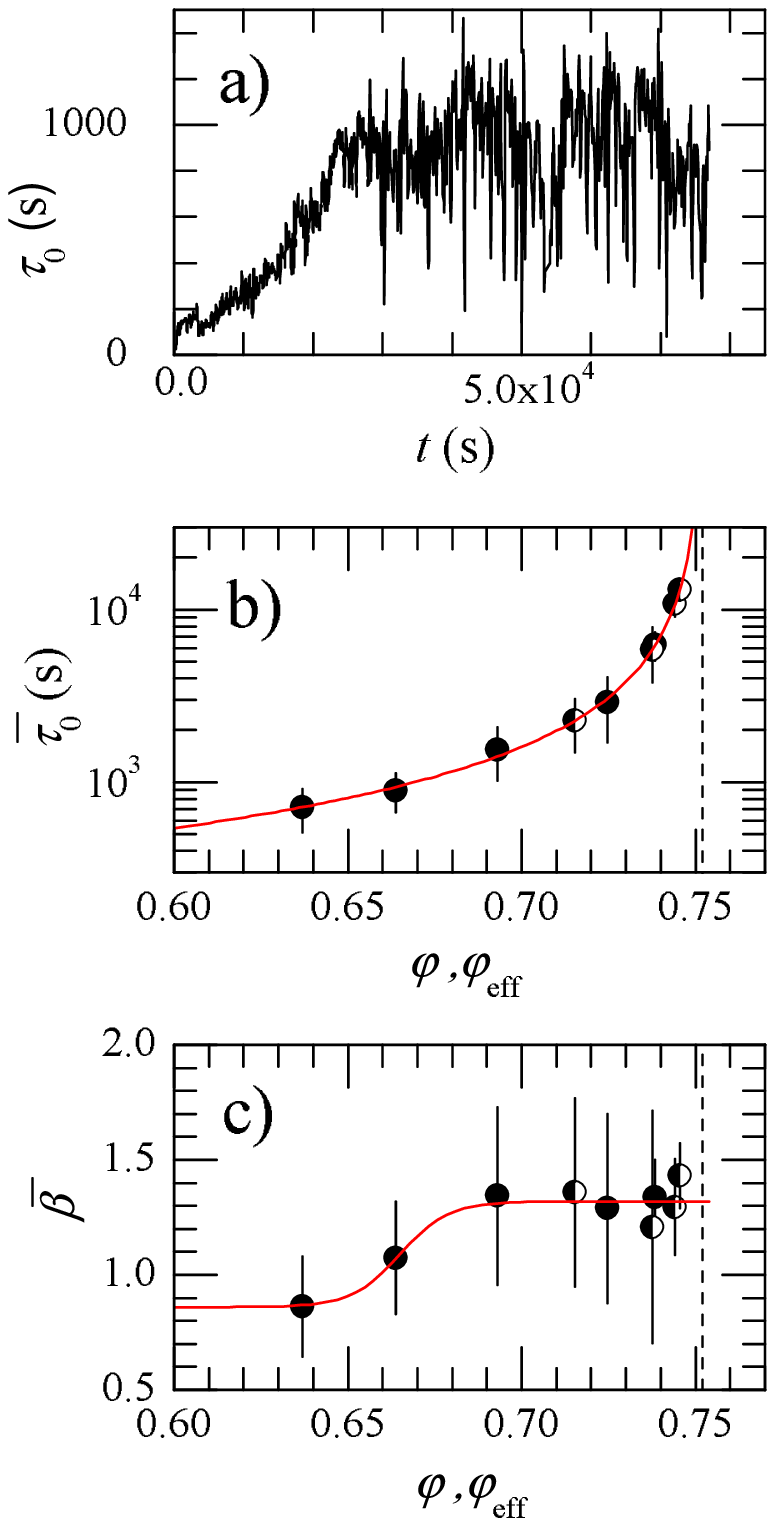}
 \caption{\textbf{Time and volume fraction dependence of the relaxation of concentrated suspensions.} a) Time dependence of the final relaxation time obtained by fitting the two-time intensity
  autocorrelation function by a stretched exponential, $g_2(t,\tau) = a(t)\exp[-(\tau/\tau_0(t))^{\beta(t)}]$, for a sample at $\varphi =
  0.664$. After an initial aging regime where the dynamics slows
  down, the system reaches a dynamically heterogeneous stationary state, where $\tau_0$
  fluctuates significantly without any overall growing trend (the time origin
is taken at the end of the sample loading and outgassing). The
lower panels
  show the volume fraction dependence of the relaxation time (b) and the stretching exponent (c) averaged
  over time in the stationary regime. The solid symbols refer to freshly
  prepared samples, the semi-filled circles to old samples whose
  dynamics has been re-initialized. The bars are the standard
  deviations of the distributions over time of $\tau_0$ and $\beta$ in the
  stationary regime. The solid line in b) is a critical law fit
  to the growth of $\tau_0$ for the fresh samples, yielding a critical exponent $x = 1.01 \pm 0.04$
   and a critical packing fraction $\varphi_{\rm max} =
  0.752$, indicated by the dashed line here and in c). The solid line
  in c) is a guide to the eyes.}
 \label{fig1xeno}
\end{figure}

\begin{figure}
\includegraphics[scale=0.8]{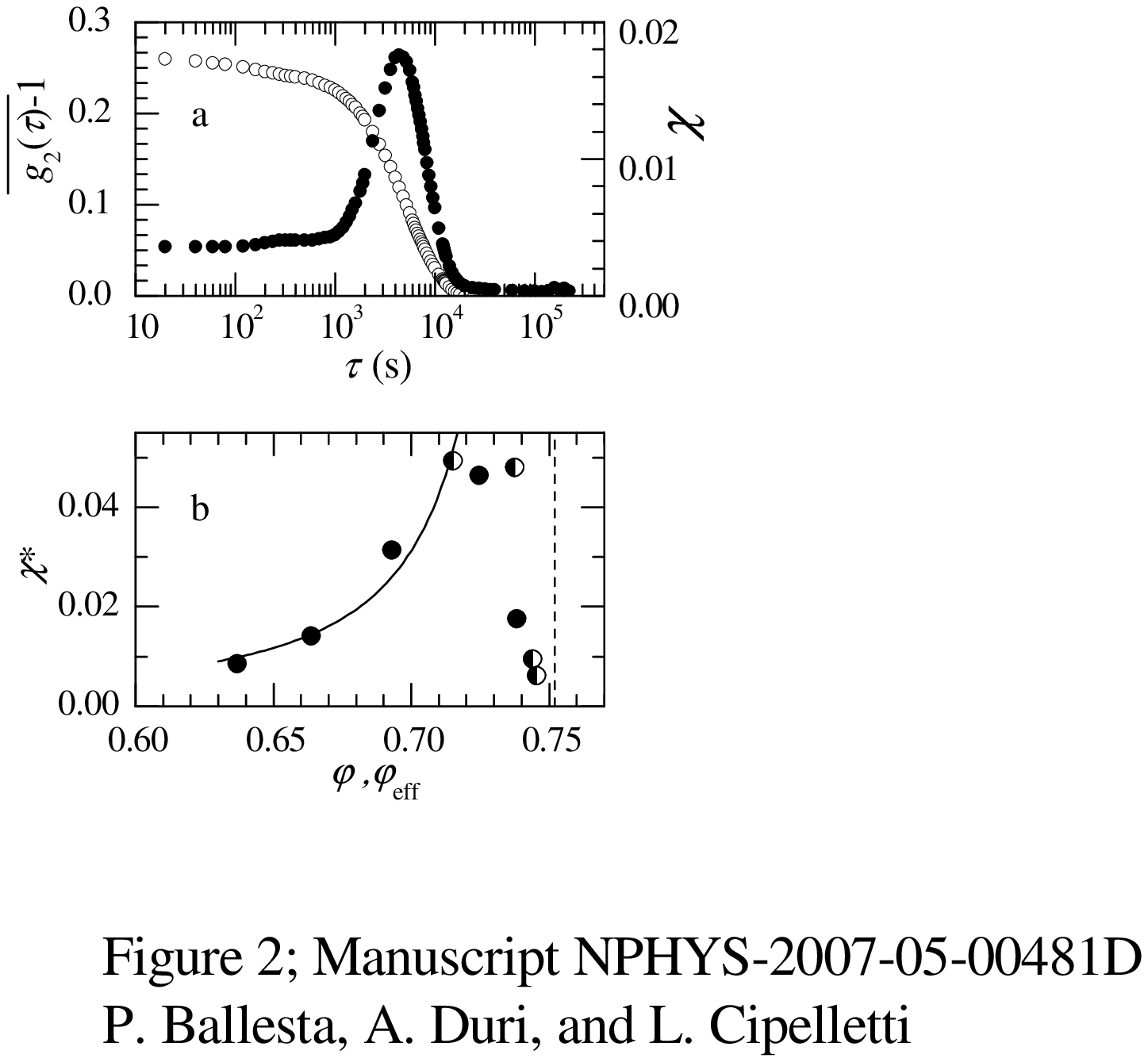}
 \caption{\textbf{Dynamical susceptibility.} a) Average correlation function $\overline{g_2(\tau)}-1$ (open symbols and left axis) and dynamical susceptibility (solid symbols and right axis),
 for a sample at $\varphi = 0.738$. b) Volume fraction dependence of the height of the peak of the dynamical
 susceptibility (same symbols as in Figs. 1b-c). The solid line is a critical-law fit to the initial
 growth of $\chi^*$, yielding an exponent $y= 1.5 \pm0.2$. Note the
 unexpected drop of $\chi^*$ near the maximum packing fraction, shown by the dashed line.}
 \label{fig2xeno}
\end{figure}

\begin{figure}
\includegraphics[scale=0.8]{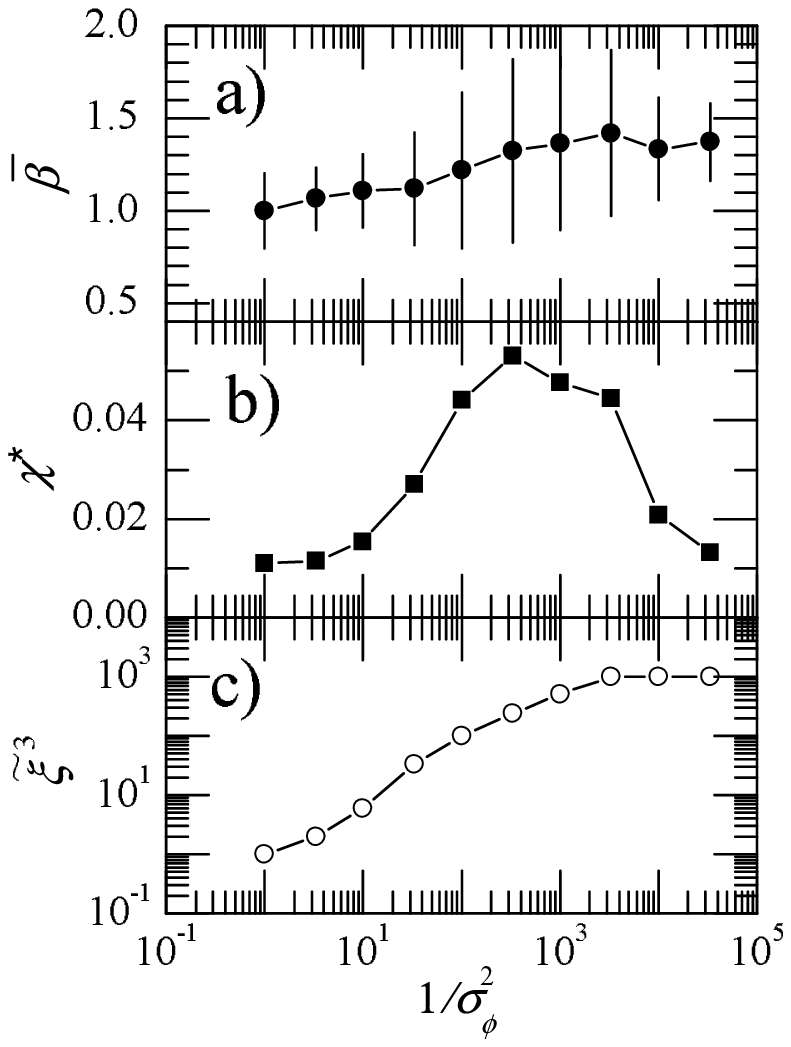}
 \caption{\textbf{Simulations.} Time-averaged stretching exponent $\overline{\beta}$ (a) and peak of the dynamical susceptibility $\chi^*$ (b)
 as a function of $1/\sigma^2_{\phi}$, as obtained from
 the simulations described in the text (note that $1/\sigma^2_{\phi}$ increases with
 $\varphi$). In a), the bars indicate the standard deviation of the
 distribution of stretching exponents.
 The volume $\widetilde{\xi}^{\,3} \equiv (\xi/\ell^*)^3$ of the rearranged
 regions used as an input in the simulations in order to reproduce the $\varphi$ dependence of $\overline{\beta}$ and $\chi^*$ is shown in (c). $\xi$
 saturates at the value of the smallest dimension of the scattering
 cell.}
 \label{fig3xeno}
\end{figure}


\newpage

\begin{references}



\bibitem{Donth} Donth, E. The Glass Transition (Springer, Berlin, 2001).

\bibitem{LiuNature1998} Liu, A. J. \& Nagel, S. R. Jamming is not Just Cool Anymore. Nature 396, 21 (1998).

\bibitem{WeeksScience2000} Weeks, E. R., Crocker, J. C., Levitt, A. C., Schofield, A. and Weitz, D. A. Three-dimensional direct imaging of structural relaxation near the colloidal glass transition. Science 287, 627-631 (2000).

\bibitem{DauchotPRL2005} Dauchot, O., Marty, G. and Biroli, G. Dynamical heterogeneity close to the jamming transition in a sheared granular material. Physical Review Letters 95, 265701 (2005).

\bibitem{Glotzer} Glotzer, S. C. Spatially heterogeneous dynamics in liquids: insight from simulation. Journal of Non-Crystalline Solids 274, 342-355 (2000).

\bibitem{Ediger} Ediger, M. D. Spatially heterogeneous dynamics in supercooled liquids. Annu. Rev. Phys. Chem. 51, 99-128 (2000).

\bibitem{BerthierScience2005} Berthier, L. et al. Direct experimental evidence of a growing length scale accompanying the glass transition. Science 310, 1797-1800 (2005).

\bibitem{KeysNaturePhys2007} Keys, A. S., Abate, A. R., Glotzer, S. C. and Durian, D. J. Measurement of growing dynamical length scales and prediction of the jamming transition in a granular material. Nature Physics 3, 260-264 (2007).


\bibitem{LacevicPRE} Lacevi\v{c}, N., Starr, F. W., Schroder, T. B., Novikov, V. N. \& Glotzer, S. C. Growing correlation length on cooling below the onset of caging in a simulated glass-forming liquid. Physical Review E 66, 030101 (2002).

\bibitem{Biroli} Biroli G., Bouchaud, J. P., Miyazaki K. \&
Reichman, D. R. Inhomogeneous mode-coupling theory and growing
dynamic length scale in supercooled liquids. Phys. Rev. Lett. 97,
195701 (2006)

\bibitem{somerefssimul} Franz, S., Donati, C., Parisi, G. \& Glotzer, S. C. On dynamical correlations in supercooled liquids. Philosophical Magazine B 79, 1827-1831 (1999).

\bibitem{JP} Chandler, D., Garrahan, J. P., Jack, R. L., Maibaum,
L. \& Pan, A. C. Lengthscale dependence of dynamic four-point
susceptibilities in glass formers. Phys. rev. E 74, 051501 (2006).

\bibitem{someREfXenospheres} Herk, H. H., Bikoles, N. M., Overgerger, C. G. \& Menzes, G. in Encyclopedy of polymer science and engineering (ed. al., H. E. M. e.) (Wiley-Interscience, New York, 2003).

\bibitem{DWSGeneral} Weitz, D. A. \& Pine, D. J. in Dynamic Light scattering (ed. Brown, W.) 652-720 (Clarendon Press, Oxford, 1993).


\bibitem{DuriPRE2005} Duri, A., Bissig, H., Trappe, V. \& Cipelletti, L. Time-resolved-correlation measurements of temporally heterogeneous dynamics. Physical Review E 72, 051401-17 (2005).

\bibitem{noteduration} Measurements in the stationary regime last
50 to 100 times longer than the average relaxation time.

\bibitem{packing} Kansal, A. R., Torquato, S. \& Stillinger, F. H. Computer generation of dense polydisperse sphere packings. Journal of Chemical Physics 117, 8212-8218 (2002).

\bibitem{LucaPRL} Cipelletti, L., Manley, S., Ball, R. C. \& Weitz, D. A. Universal aging features in the restructuring of fractal colloidal gels. Physical Review Letters 84, 2275-2278 (2000).

\bibitem{LaponiteXPCS} Bandyopadhyay, R. et al. Evolution of particle-scale dynamics in an aging clay suspension. Physical Review Letters 93, 228302 (2004).

\bibitem{ReviewLeheney} Bandyopadhyay, R., Liang, D., Harden, J. L. \& Leheny, R. L. Slow dynamics, aging, and glassy rheology in soft and living matter. Solid State Communications 139, 589-598 (2006).

\bibitem{Munch} Knaebel, A. et al. Aging behavior of laponite clay particle suspensions. Europhys. Lett. 52, 73 (2000).

\bibitem{Leheny} Chung, B. et al. Microscopic dynamics of recovery in sheared depletion gels. Physical Review Letters 96, - (2006).

\bibitem{notenorm} In simulations, $\chi_4$ is normalized by multiplying the variance of the intermediate scattering
function (ISF) ---or a similarly-defined correlation function--- by
the number of particles. Therefore, to compare the order of
magnitude of $\chi$ to that of $\chi_4$ one should multiply the
former by the number of particles in the scattering volume, $N_p$,
and take the square root, since $g_2-1$
is homogeneous to a squared ISF. For the experiments reported here,
the conversion factor is of order $\sqrt{N_p} \approx 5 \times
10^3$.

\bibitem{MayerPRL2004} Mayer, P. et al. Heterogeneous Dynamics of Coarsening Systems. Physical Review Letters 93, 115701 (2004).

\bibitem{Weekscage} Weeks, E. R. \& Weitz, D. A. Properties of cage rearrangements observed near the colloidal glass transition. Physical Review Letters 89 (2002).

\bibitem{DauchotCage} Marty, G. \& Dauchot, O. Subdiffusion and cage effect in a sheared granular material. Physical Review Letters 94, 015701 (2005).

\bibitem{PRLRecentGranular} Reis, P. M., Ingale, R. A. \& Shattuck, M. D. Caging Dynamics in a Granular Fluid. Physical Review Letters 98, 188301 (2007).

\bibitem{Wyart} Brito, C. \& Wyart, M. Heterogeneous Dynamics, Marginal Stability and Soft Modes in Hard Sphere Glasses. cond-mat/0611097 (2007).

\bibitem{DurianScience1992} Durian, D. J., Pine, D. J. \& Weitz, D. A. Multiple light-scattering probes of foam structure and dynamics. Science 252, 686-688 (1991).

\bibitem{DuriEPL2006} Duri, A. \& Cipelletti, L. Length scale dependence of dynamical heterogeneity
in a colloidal fractal gel. Europhysics Letters 76, 972-978 (2006).

\bibitem{notemsd} Following ref.~\cite{DWSGeneral}, one finds
$\sigma^2_{\phi} = 2k_0^{2}\sigma^2\ell/(3\ell^*)$, where $k_0$ is
the wave vector of the incident light and $\ell$ is the photon
scattering mean free path. For particles much larger than the laser
wavelength, as in our case, $\ell \approx 0.1\ell^*$, yielding
$\sigma^2_{\phi} \approx  20 \sigma^2/\mu{\mathrm m}^2$.

\bibitem{noterate} An additional input parameter is the
rearrangement rate per unit volume. However, this parameter only
sets the time scale for the simulations with no impact on
$\overline{\beta}$ or $\chi^*$

\bibitem{notesaturation} Using a steepest descent method, one
can show that for $1/\sigma^2_{\phi} \rightarrow \infty$
$\overline{g_1^{(s)}(\tau)}$ has a compressed exponential shape
with a compressing exponent equal to $p$. Summing over all paths
with different length $s$ --and thus different decay rates--
results in an effective compressing exponent $\overline{\beta} <
p$.

\bibitem{Kob2000} Kob, W. \& Barrat, J. L. Fluctuations, response and aging dynamics in a simple glass-forming liquid out of equilibrium. European Physical Journal B 13, 319
(2000).

\bibitem{ReviewConfinement} Alcoutlabi, M. \& McKenna, G. B. Effects of confinement on material behaviour at the nanometre size scale. Journal of Physics-Condensed Matter 17, R461-R524 (2005).

\end{references}
\end{document}